\documentclass[journal=jacsat,manuscript=article]{achemso}
\usepackage{longtable}
\usepackage{multirow}
\usepackage{amsmath}
\usepackage{subfigure}
\usepackage[usenames,dvipsnames]{color}
\SectionNumbersOn
\author{Marcos H. Gim\'{e}nez}
\affiliation{Departamento de F\'{\i}sica Aplicada, Universitat
Polit\`{e}cnica de Val\`{e}ncia, Cam\'{\i} de Vera s/n, 46022,
Val\`{e}ncia, Spain.}
\author{Juan Carlos Castro-Palacio}
\affiliation{Department of Earth Sciences and Engineering, Faculty
of Engineering, Imperial College London, UK.}
\author{Jos\'{e} Antonio G\'{o}mez-Tejedor}
\affiliation{Centro de Biomateriales e Ingenier\'{\i}a Tisular,
Universitat Polit\`{e}cnica de Val\`{e}ncia, Cam\'{\i} de Vera s/n,
46022, Val\`{e}ncia, Spain.}
\author{Luisberis Velazquez}
\affiliation{Departamento de F\'{\i}sica, Universidad Cat\'{o}lica
del Norte. Av. Angamos 0610, Antofagasta, Chile.}
\author{Juan A. Monsoriu}
\email{jmonsori@fis.upv.es} \affiliation{Centro de Tecnolog\'{\i}as
F\'{\i}sicas, Universitat Polit\`{e}cnica de Val\`{e}ncia, Cam\'{\i}
de Vera s/n, 46022, Val\`{e}ncia, Spain.}
\date{\today}

\title{Theoretical and experimental study of the normal modes in a coupled two-dimensional system}

\begin{document}

\begin{abstract}
In this work, the normal modes of a two-dimensional oscillating
system have been studied from a theoretical and experimental point
of view. The normal frequencies predicted by the Hessian matrix for
a coupled two-dimensional particle system are compared to those
obtained for a real system consisting of two oscillating smartphones
coupled one to the other by springs. Experiments are performed on an
air table in order to remove the friction forces. The oscillation
data are captured by the acceleration sensor of the smartphones and
exported to file for further analysis. The experimental frequencies
compare reasonably well with the theoretical predictions,
specifically, within 1.7\% of discrepancy.

PACS: 46.40.-f; 43.20.+g; 43.40.+s

\end{abstract}

\section{Introduction}

 \noindent
The study of the normal modes is a central issue in understanding
the properties of solids and molecules, such as solid phonons and
vibrations of polyatomic
molecules.\cite{sonia,jccp_modes1,jccp_modes2,lomb}. Therein, the
formalism of the Hessian matrix is a common approach.\cite{ash}
Therefore, this topic is included in the courses of Physics and
Chemistry degrees higher in the syllabus. For instance, the
collective oscillations of a periodic solid, the phonons, which
reveal important information, e.g. about thermal and electrical
conductivity can be derived experimentally from neutron scattering.
In the case of polyatomic molecules, normal modes are connected to
the vibrational spectrum, which can be measured using a number of
spectroscopic techniques. From a pedagogical point of view, the
simplest classical model to characterize the vibrational modes of a
polyatomic molecule is a particle system coupled by pair
potentials.\cite{wilson}

 \noindent
In general physics courses, the topic of coupled systems has been
basically analyzed by means of linear 1D models.\cite{resnick} It is
also possible to find a number of works in the literature on the
experimental characterization of coupled 1D systems connected to
external drivers \cite{ryan}, i.e. by using video-analysis
techniques\cite{monso}, electromechanical systems\cite{molina} or
sensors.\cite{jua2} However, when it comes to everyday life, most
oscillations are more than one-dimensional. This is a good reason
for including two-dimensional oscillation experiments in physics
teaching.\cite{bob,huerta}

 \noindent
Simple experiments involving oscillations are largely facilitated by
introducing smartphones as oscillating bodies in one\cite{vog4} and
two dimensions.\cite{luis} The acceleration sensor carried by these
devices can be used to collect the oscillation data which can be
exported to file for further analysis.\cite{jua1} This is a major
advantage since the way of studying two-dimensional oscillations in
previous work\cite{bob} was somewhat tedious. For example, the
trajectory of an oscillating puck on an air table can be followed by
the trace and described by it onto paper, which is later digitalized
to extract the information of the trajectory\cite{bob}. The
introduction of the smartphone acceleration sensor to measuring this
kind of two dimensional oscillations represented a major progress in
our previous work\cite{luis} where mechanical Lissajous figures were
obtained in a very simple way.

 \noindent
In this work, we present an exhaustive theoretical and experimental
study of the normal modes in a coupled 2D system. The experimental
setup consists of two smartphones on an air table connected each
other by springs and to fixed ends. The air table allows us to
remove the friction forces. In these experiments the mobile phones
themselves are the bodies under study. The coupled oscillations are
captured with the acceleration sensors of the smartphones and the
data are exported to file for further analysis.

 \noindent
It should be pointed out that the smartphones are just measurement
tools here. Its use is not the main contribution of this work. In
fact, two-dimensional oscillations could be also analyzed by using
other techniques, i.e. video analysis techniques.\cite{monso,hueso}.
However we have preferred to use smartphones since they allow for a
fast and direct acquisition of data. Based on the collected data,
the normal modes in the 2D system of coupled oscillators can be
deeply analyzed, which is the main objective of this work. The
theoretical frequencies derived from this analysis based on the
Hessian matrix are compared with those obtained from processing the
smartphone sensor data. In this way, we provide an example of
physics teaching experiment on 2D coupled oscillations which
contributes to fill the existing gap in the General Physics courses.
In this simple way, students may be introduced to the vibrational
properties of solids and molecules.

 \noindent
The outline of the paper is the following. In section 2, we describe
the calculation of the normal modes from the Hessian matrix
formalism applied to a coupled two-dimensional particle system. In
section 3, experiments using two smartphones as oscillating bodies
on an air table are described. The processing of the oscillation
data and the comparison between the experimental and calculated
normal frequencies are then presented. Finally, in section 4, some
conclusions are drawn.

\section{Hessian matrix formalism}
\begin{figure}[H]
\centering
\includegraphics[scale=0.60]{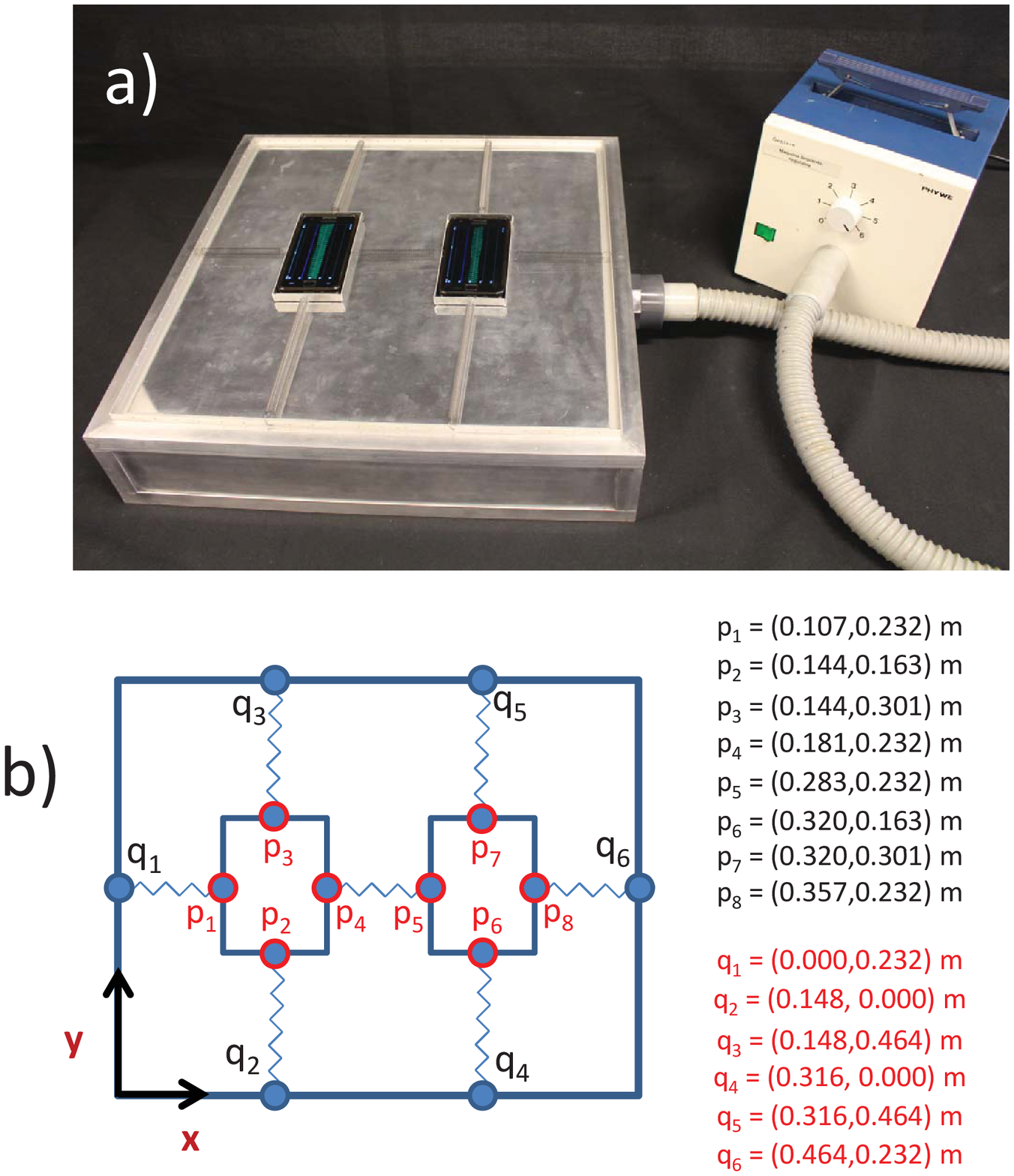}
\caption{Photograph of the experimental setup. In panel a), a global
view of the squared air table, the air supplier, the springs, and
the smartphones are shown. The geometric parameters of the system at
rest are shown in panel b).} \label{fig:fig1}
\end{figure}

A photograph of the experimental setup used for obtaining the
vibrational normal modes in a coupled 2D systems is shown in Figure
\ref{fig:fig1}a. It consists of the air table, the air supplier, the
springs, and two smartphones Samsung Galaxy S2 GT-I9100 bearing an
Android version 4.03. The mass of the smartphones (plus the carrying
tray) is $m$=(174.4 $\pm$ 0.1) g for both smartphones. As indicated
in the figure, the lay out of the spring is a two-plus-signs
geometry. The air table is a square of side (0.464$ \pm$ 0.001) m.
The force constant of the springs is $k$=(20.6$ \pm$ 0.1) N/m and
its natural length  is $d$=(0.058 $\pm $0.001) m. The remaining
geometric parameters of the system are shown in Figure
\ref{fig:fig1}b.

 \noindent
First of all, the normal frequencies for the coupled system of
figure 1 are calculated by a methodology based on the Hessian
matrix. In this respect, the total potential energy of the system
can be calculated taking into account the geometric variables
defined in Figure \ref{fig:fig1}b and the displacement vectors,
\begin{equation}
\begin{split}
\Delta \overrightarrow{r_1} = \overrightarrow{r_1} - \overrightarrow{r_{1,0}} = (x_1-x_{1,0},y_1-y_{1,0}), \\
 \Delta \overrightarrow{r_2}= \overrightarrow{r_2} - \overrightarrow{r_{2,0}} =
 (x_2-x_{2,0},y_2-y_{2,0}),
  \label{disp}
\end{split}
\end{equation}
where $\overrightarrow{r_{1,0}} = (x_{1,0},y_{1,0} )$ and
$\overrightarrow{r_{2,0}} = (x_{2,0},y_{2,0} )$ are the vector
positions of the smartphones at the equilibrium positions and
$\overrightarrow{r_1} = (x_1,y_1)$ and $\overrightarrow{r_2} =
(x_2,y_2)$ are the corresponding vector positions when the
smartphones are in motion.

 \noindent
It should be noted that the springs stretch approximately three
times their natural length. In this respect, we have made an
independent experiment to check the linearity of the springs. In
these conditions and considering the harmonic approximation, the
total potential is given by,
\begin{equation}
U=\frac{k}{2}\overset{7}{\underset{i=1}{\sum}}(d_{i}-d_{0})^{2},
\label{etot1}
\end{equation}
where the elongation of each spring, $d_{i}$, can be determined from
the points represented in Figure \ref{fig:fig1}b and from the
displacements in Eq. \ref{disp},
\begin{equation}
\begin{split}
 d_1 = |\overrightarrow{q_1p_1} + \Delta \overrightarrow{r_1}|, \\
 d_2 = |\overrightarrow{q_2p_2} + \Delta \overrightarrow{r_1}|, \\
  d_3 = |\overrightarrow{q_3p_3} + \Delta \overrightarrow{r_1}|, \\
 d_4 = |\overrightarrow{q_4p_6} + \Delta \overrightarrow{r_2}|, \\
  d_5 = |\overrightarrow{q_5p_7} + \Delta \overrightarrow{r_2}|, \\
 d_6 = |\overrightarrow{q_6p_8} + \Delta \overrightarrow{r_2}|, \\
  d_7 = |\overrightarrow{p_4p_5} + \Delta \overrightarrow{r_2} - \Delta \overrightarrow{r_1}|. \\
\end{split}
\label{distance}
\end{equation}

 \noindent
It should be noticed that $d_{0}$ is the natural length of the
spring. Thus, at rest ($\Delta \overrightarrow{r_1}=\Delta
\overrightarrow{r_2}=(0,0)$), the seven springs are elongated and
the energy of the system represented by Eq.(\ref{etot1}) is minimal
but not zero.

 \noindent
From Figure \ref{fig:fig1}, it appears that there are 6 degrees of
freedom, three for the center of mass of each smartphones and
$\theta$ for the rotation about the center of mass of the system.
However, we have not considered rotations in our two-dimensional
model consisting of two coupled particles. Under these conditions,
and taking into account that oscillations take place on the x,y
plane, we have a system with four degrees of freedom, namely,
translations along x- and y- axes for each smartphone. The dynamical
matrix (Hessian matrix)\cite{ash} is then expressed as,
\begin{equation}
D_{i_{\alpha}j_{\beta}}=\frac{1}{m}\big(\frac{\partial^{2}U}{\partial
u_{i_{\alpha}}\partial
u_{j_{\beta}}}\big)_{u_{i_{\alpha}}=0;u_{j_{\beta}}=0},
\label{din}%
\end{equation}
where $U$ is the total potential energy and $u_{i_{\alpha}}$
($u_{j_{\beta}}$) is the displacement of the \textit{i}-th or
\textit{j}-th particle ($i,j=1,2$) along the $\alpha$ or $\beta$
axis ($\alpha,\beta=x,y$).

\noindent   The evaluation of this matrix at the equilibrium
positions and further diagonalization yields the four vibrational
eigenfrequencies squared. These normal frequencies will be denoted
as $\omega_{x}^{S}$, $\omega_{y}^{S}$, $\omega_{x}^{A}$, and $\omega
_{y}^{A}$, corresponding to the symmetric and antisymmetric modes
and for the x- and y- axes, respectively.

\noindent By using this potential energy expression,
$U=U(x_1,y_1,x_2,y_2)$ given by Eq.(\ref{etot1}), the above Hessian
matrix, evaluated at the equilibrium positions is,
\begin{equation}
D=\left(
\begin{tabular}
[c]{llll}%
388.493 & 0.000 & -118.119 & 0.0  \\
0.000 & 341.233 &
0.000 & -50.953 \\
-118.119 & 0.000 & 388.493 &
0.000 \\
0.000 & -50.953 & 0.000 &
341.233%
\end{tabular}
\right)
\end{equation}

 \noindent
The resulting normal modes (the square root of the eigenvalues) are,
$\omega_{x}^{S}=16.443$ rad/s, $\omega_{y}^{S}=17.038$ rad/s,
$\omega_{x}^{A}=22.508$ rad/s, and $\omega _{y}^{A}=19.804$ rad/s.
It should be noted that these values are only valid for small
displacements about the equilibrium positions. It can also be noted
that the eigenfrequencies obtained for the x- axis are significantly
different from the ones from the 1D model, that is:
$(\omega_{x}^{A})^2 = 3 (\omega_{x}^{S})^2$. This is due to the
effect of the vertical springs (p2q2, p3q3, p6q4 and p7q5) on the
horizontal oscillations (for us, ``horizontal" is when the
oscillation is along the x- axis and ``vertical" along the y- axis).
In addition, and as expected, the eigenfrequencies along the y- axis
also differ from the ones obtained for the x- axis. For instance,
the symmetric mode along the y- axis is affected by the horizontal
springs, namely,  p1q1 and p8q6 (p4p5 does not stretch in this
case), while the symmetric mode along the x- axis is affected by the
four springs aforementioned. However, in the case of the
antisymmetric mode along the y- axis, the horizontal mode p4p5 is
affected.

 \noindent
It is also possible to perform a more exhaustive study of the normal
modes of oscillation by using Newton's second law. For example, as
for the horizontal symmetric mode, the total potential given by
Eq.(\ref{etot1}) can be particularized as, $\Delta
\overrightarrow{r_1} = \Delta \overrightarrow{r_2} = (\Delta x,0)$
where $\Delta x = x-x_0$ is a synchronized displacement of both
bodies along the x- axis direction. In this situation, the potential
energy only depends on the global displacement $\Delta x$. Taking
into account that all involved forces are conservative, the net
force acting on the system is,
$\overrightarrow{F}=-\overrightarrow{\bigtriangledown}U$. For the
particular case under consideration $F(x)=-dU/dx$. On the other
hand, Newton's second law can be expressed as
$F(x)=m_{Total}\frac{d^2 x}{dt^2}$, where $m_{Total}=2m$ is the
total mass of the system. Therefore, the resulting nonlinear
differential equation governing the system is,

\begin{equation}
-\frac{dU}{dx}=2m\frac{d^2 x}{dt^2}.
 \label{segNew}
\end{equation}

 \noindent
It should be pointed out that in Eq. \ref{segNew}, which governs the
symmetric horizontal mode, elastic forces of both horizontal and
vertical springs are present. The vertical springs stretch even when
the particles move along the \textit{x} axis only. Contrary to a
simple 1D model of coupled oscillations, there is no analytical
solution of Newton's second law for the 2D case and so a numerical
solution is required.

\begin{figure}[H]
\centering
\includegraphics[scale=0.40]{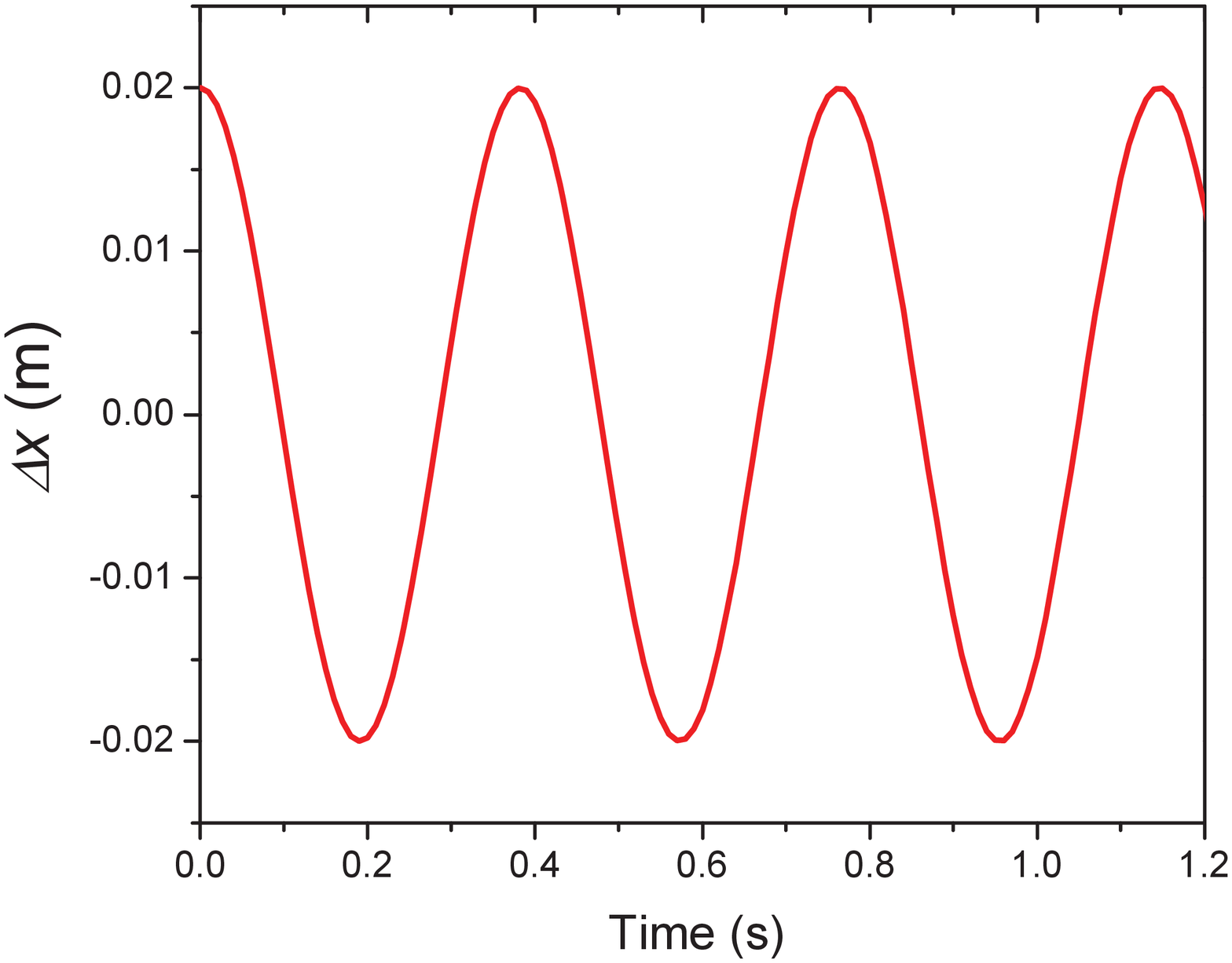}
\caption{Simultaneous displacement of the bodies along the x axis
\textit{versus} time. The curves of the numerical (exact) and
harmonic trajectories overlap visually.} \label{fig:fig2}
\end{figure}

 \noindent
 By using the function NDSolve of the software
Mathematica, Eq.\ref{segNew} can be solved numerically using $\Delta
x_0$ and $\frac{dx}{dt}=0$ as initial conditions. For an initial
displacement $\Delta x_0=0.02$ m, the numerical solution of Eq.
\ref{segNew} provides the trajectory displayed in Figure
\ref{fig:fig2} (solid line). Additionally, the harmonic oscillation
$\Delta x(t)=Acos (\omega^S_x t)$ with $A=\Delta x_0$ is shown in
the same figure (dashed line), but it can not be seen since the
curves overlap visually.

 \noindent
From Figure \ref{fig:fig2}, the exact period of oscillations can be
determined, $T=0.3818$ s, and from it the exact (from solving
Eq.\ref{segNew}) value of the frequency, $\omega =16.457$ rad/s. The
discrepancy between this value and the harmonic result
($\omega_{x}^{S}=16.443$) is only $0.09\%$. The small discrepancy
between both results is due to the influence of the vertical
springs. The horizontal projections of the forces exerted by these
springs is linear only for small displacements. This study can be
repeated for the antisymmetric mode by imposing $\Delta
\overrightarrow{r_1} = - \Delta \overrightarrow{r_2}$ in the
potential $U$.

\noindent A similar analysis for the symmetric and antisymmetric
normal modes along the y- axis, and using $0.02$ m as initial
displacement, yields discrepancies between the harmonic and the
exact frequencies within 1 $\%$ in all cases. The smaller the
initial displacement the smaller the discrepancy. For instance,
discrepancies within 0.3 $\%$ are obtained if 1 cm is used as
initial displacement. The smaller the displacements the better the
harmonic approximation approaches the physical experiment. Thereby,
the Hessian matrix formalism constitutes a very good approximation
for obtaining the normal frequencies of a coupled 2D system in basic
Physics courses.

\section{Experimental setup}

 \noindent
In order to check the normal frequencies predicted from the Hessian
matrix, experiments using the experimental setup of Figure
\ref{fig:fig1} are carried out. The oscillation data are captured by
the acceleration sensor of the smartphones. From previous
experiments, we already know that the acceleration sensor in our
smartphone's models is located at the center of the smartphone,
which is coincident with the center of mass of the
system.\cite{jua3}. However, the position of the acceleration sensor
may not be at the geometrical center for other models.\cite{mau}

 \noindent
For the interaction with the mobile sensor, the free Android
application ``Accelerometer Toy ver 1.0.10" is used. This
application takes 316 kB of SD card memory and can be downloaded
from the Google play website. \cite{ac-mon}. The values of the
acceleration components on $x,y$ and $z$ - axes are registered at
each time step. The precision in the measurement of the acceleration
is $\delta a$ = 0.03 m/s$^{2}$ and of time is $\delta t$ = 0.01 s.
This application also allows to save the output data to file from
which further analysis can be performed. Once the application is
downloaded to the mobile device, a small test can be done to ensure
the device is working correctly. If the mobile is left undisturbed
on a horizontal surface, the application output curves for the
acceleration should indicate values very close to zero for all axes.
This application was successfully used in other experiments to study
uniform and uniformly accelerated circular motions.\cite{jua3}

 \noindent
 Five experiments are performed using the setup of Figure
\ref{fig:fig1}. In the first four experiments, the system is set to
oscillate by hand with approximately normal frequencies (symmetric
and antisymmetric) along x- and y-axes, respectively. For the case
of the symmetric mode, mobile phones are displaced about 1 cm
towards the positive x-axis and towards the positive y- axis,
respectively. For the antisymmetric mode, one of the mobile phones
is displaced to the left and the other to the right for the x-axis,
and downward and upward for the y-axis, respectively.

\begin{figure}[H]
\centering
\includegraphics[scale=0.60]{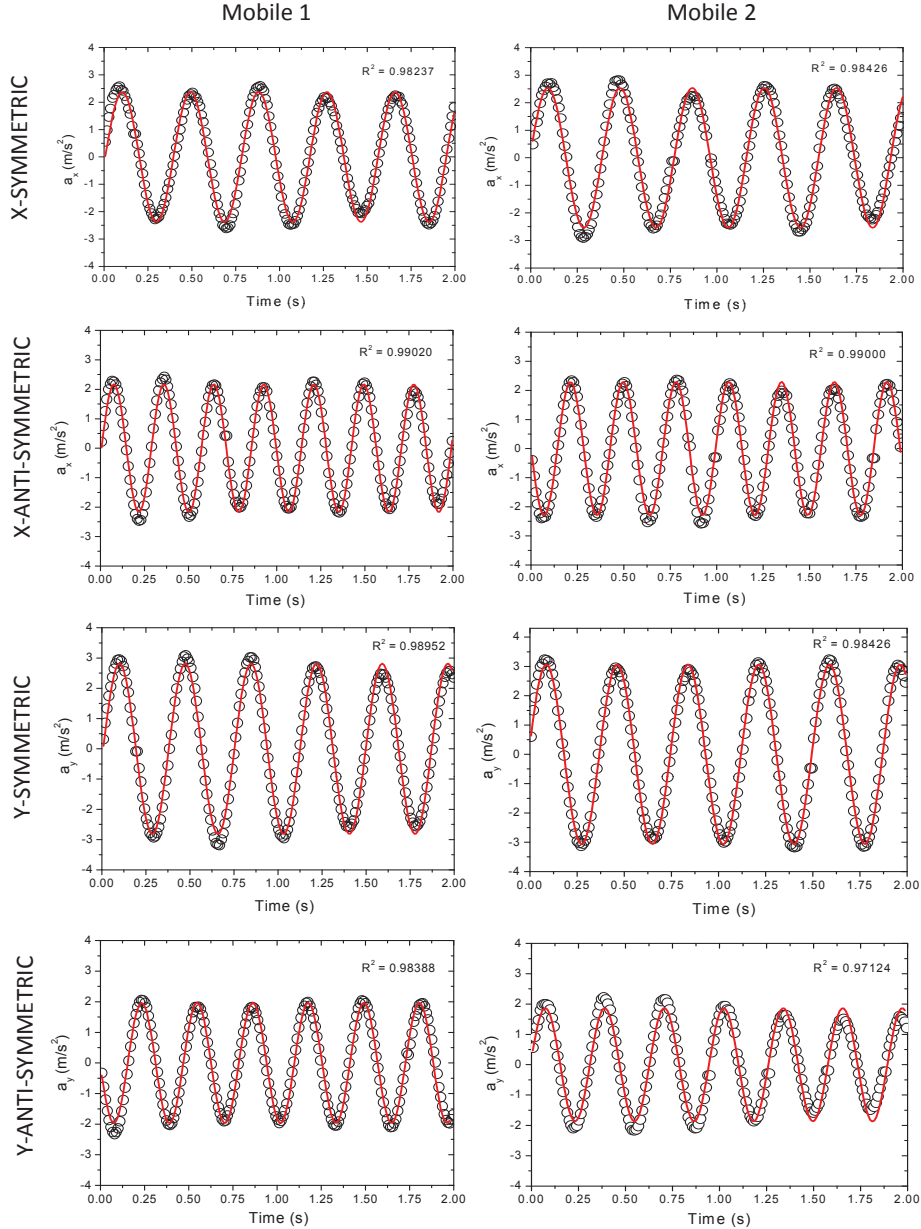}
\caption{Symmetric and antisymmetric acceleration oscillations of
the mobiles 1 and 2 along x- and y- axes respectively. The red solid
lines indicate the fit. The square of the curvilinear correlation
coefficient, $R^2$, has been included in the graphs.}
\label{fig:fig3}
\end{figure}

 \noindent
The data registered by the acceleration sensor of each smartphone
for the symmetric and antisymmetric oscillations (see Figure 3) can
be fitted to a harmonic function, $a(t)=A\sin(\omega t+\phi)$ where
$A$ is the amplitude, $\omega$ the frequency and $\phi$ the phase.
The results for the frequencies are registered in Table I. There are
8 cases in total, that is, considering each axis and each
smartphone, for the symmetric and anti-symmetric modes,
respectively. The graphs of the acceleration measurements and the
corresponding fit curve are included in Figure \ref{fig:fig3} for
each smartphone, normal mode and axis.

 \noindent
{Table I. Frequencies and uncertainties from the fit of the
acceleration data to $a(t)=A\sin(\omega t+\phi)$ along x- and y-
axes for the mobiles 1 and 2, respectively.}
\[%
\begin{tabular}
[c]{ccc}\hline\hline   & Mobile 1 & Mobile 2 \\\hline
 $\omega_{x}^{S}$ (rad/s)  & 16.158$ \pm $0.016 & 16.207$ \pm $0.016 \\
 $\omega_{y}^{S}$ (rad/s)  &16.854$ \pm $0.013 & 16.732$ \pm $0.010 \\
 $\omega_{x}^{A}$ (rad/s)  & 22.158$ \pm $0.012 & 22.115$ \pm $0.012 \\
 $\omega_{y}^{A}$ (rad/s)  & 19.988$ \pm $0.017 & 19.860$ \pm$ 0.020 \\\hline
\end{tabular}
\label{tab1}
\]

 \noindent
{Table II. Comparison between the experimental results (average
values from Table I) and those obtained from the Hessian matrix
formalism.}
\[%
\begin{tabular}
[c]{cccc}\hline\hline   & Experimental results & Hessian matrix
formalism & Discrepancies (\%)
\\\hline
 $\omega_{x}^{S}$ (rad/s)  & 16.18 $\pm$ 0.03 & 16.443  & 1.6 \\
 $\omega_{y}^{S}$ (rad/s)  & 16.79 $\pm$ 0.02 & 17.038  & 1.5 \\
 $\omega_{x}^{A}$ (rad/s)  & 22.14 $\pm$ 0.02 & 22.508  & 1.7 \\
 $\omega_{y}^{A}$ (rad/s)  & 19.92 $\pm$ 0.04 & 19.804  & 0.6 \\\hline
\end{tabular}
\label{tab2}
\]

\noindent The analogue values to the frequencies of the smartphones
for the symmetric and antisymmetric modes and for the x- and y- axes
are shown in Table II. The corresponding normal frequencies obtained
from the Hessian matrix formalism are also included. A very good
agreement is obtained between the experimental and the theoretical
results.

 \noindent
Finally, in the fifth experiment, an arbitrary oscillation is
started by just shifting one of the mobiles out of the equilibrium
position. In this case, the  arbitrary oscillation (non-normal) of
the studied system can be represented as a superposition of the
corresponding four normal oscillations,
\begin{equation}
a(t)=A_{x}^{S}\sin(\omega_{x}^{S}t+\phi_{x}^{S})+A_{x}^{A}\sin(\omega_{x}%
^{A}t+\phi_{x}^{A})+A_{y}^{S}\sin(\omega_{y}^{S}t+\phi_{y}^{S})+A_{y}^{A}%
\sin(\omega_{y}^{A}t+\phi_{y}^{A})\text{.}
 \label{eqfree}
\end{equation}

\begin{figure}[H]
\centering
\includegraphics[scale=0.50]{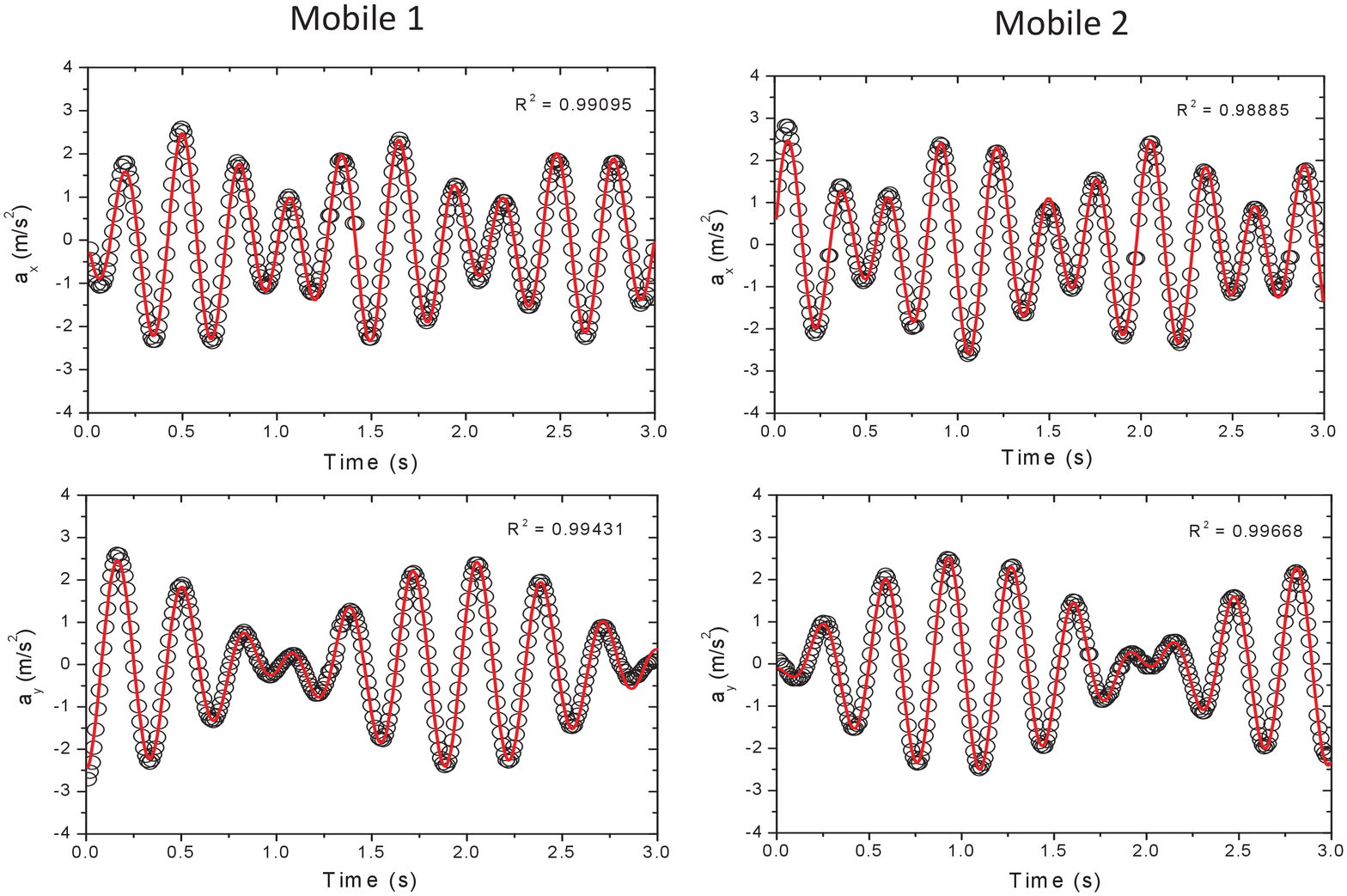}
\caption{Free oscillations of the smartphones. The red solid lines
indicate the fit.} \label{fig:fig4}
\end{figure}

\noindent Figure \ref{fig:fig4} shows the data points for an
arbitrary oscillation. The curve fit to Eq.\ref{eqfree} is indicated
with a solid red line. The fitting to Eq. \ref{eqfree} has been
carried out by using the non-linear fitting algorithm
Levenberg-Marquardt \cite{lev, mar}.

 \noindent
In all cases shown in Figure \ref{fig:fig4} the values of $R^2$ are
around 0.99 what indicates the good quality of the fitting
procedure. The corresponding fitted frequencies are not shown for
brevity, since they are very similar as those reported in Tables I
and II. Alternatively, the main frequencies of the system can be
also explored by using the Fourier transform of a free oscillation
of acceleration data. Our objective was rather to prove the validity
of the Hessian matrix formalism in predicting the normal frequencies
of a 2D coupled system. To connect basic and simple oscillation
experiments like the one in this article with this formalism helps
prepare the student's mindset for physics courses further in the
syllabus.

\section{Conclusions}
The normal frequencies of a coupled two-dimensional system are
studied both theoretically and experimentally. The normal modes were
first calculated for a particle system from the Hessian matrix. An
experimental setup using smartphones instead of particles and with
real springs is used to test the theoretical model. The oscillation
data were collected by the acceleration sensor of the smartphones.
For all cases, the percentage discrepancies between the theoretical
and experimental frequencies are within 1.7 \%.

\section*{Acknowledgments}
Authors would like to thank the Institute of Educational Sciences of
the Universitat Polit\`{e}cnica de Val\`{e}ncia (Spain) for the
support of the Teaching Innovation Groups MoMa and e-MACAF and for
the financial support through the Project PIME 2015 B18.


\begin{thebibliography}{99}

\bibitem {sonia} S. Jimenez, A. Pasquarello, R. Car and M. Chergui, Chem. Phys. 233 (1998) 343.

\bibitem {jccp_modes1} J. C. Castro-Palacios, L. Velazquez, G. Rojas-Lorenzo and J. Rubayo-Soneira, J. Mol. Struct. (Theochem) 730 (2005) 255.

\bibitem {jccp_modes2} J. C. Castro-Palacio, L. Velázquez, A. Lombardi, V. Aquilanti and J. Rubayo-Soneira, J. Chem. Phys. 11126 (2007)
174701.

\bibitem {lomb} A. Lombardi, F. Palazzetti, G. Grossi, V. Aquilanti, J. C. Castro-Palacio and J. Rubayo-Soneira, Phys. Scr. 80 (2009) 048103.

\bibitem {ash} N. W. Ashcroft and N. D. Mermin, Solid State Physics 1st edition (Saunders, Philadelphia, 1976).

\bibitem {wilson} E. B. Wilson Jr., J. C: Decius, P. C. Cross, Molecular Vibrations: The Theory of Infrared and Raman Vibrational Spectra (New York:
Dover, 1995)

\bibitem {resnick} R. Resnick, D. Halliday and K. S. Krane, Physics 4th edition (Mexico, DF: CECSA, 1999)

\bibitem {ryan} R. Givens, D. F. de Alcantara-Bonfim and R. B. Ormond, Am. J. Phys. 71 (2003) 87.

\bibitem{monso} J. A. Monsoriu, M. H. Gim\'{e}nez, J. Riera and A. Vidaurre, Eur. J. Phys. 26 (2005)
1149.

\bibitem {molina} J. E. Molina-Coronell abnd B.P. Rodr\'{i}guez-Villanueva, Rev. Mex. Fis.
E 61 (2015) 65.

\bibitem {jua2} J. C. Castro-Palacio, L. Vel\'{a}zquez-Abad, F. Gim\'{e}nez and J. A. Monsoriu, Eur. J. Phys.
34 (2013) 737.

\bibitem {bob} N. C. Bobillo-Ares and J. Fernandez-Nunez, Eur. J.
Phys. 16 (1995) 223.

\bibitem {huerta} J. S. P\'{e}rez-Huerta, C. Meneses-Fabi\'{a}n and G. Rodriguez-Zurita,
Rev. Mex. Fis. E 55 (2009) 8.

\bibitem {vog4} J. Kuhn and P. Vogt, Phys. Teach. 50 (2012) 504.

\bibitem {luis} L. Tuset-Sanchis, J. C. Castro-Palacio, J. A. G\'{o}mez-Tejedor, F. J. Manj\'{o}n and J. A. Monsoriu, Phys. Ed. 50 (2015)
580.

\bibitem {jua1} J. C. Castro-Palacio, L. Vel\'{a}zquez-Abad, M. H. Gim\'{e}nez and J. A. Monsoriu, Am. J. Phys. 81 (2013)
472.

\bibitem {hueso} J. Riera, J. A. Monsoriu, M. H. Gim\'{e}nez, J. L. Hueso and J. R. Torregrosa, Am. J. Phys. 71 (2003)
1075.

\bibitem {ac-mon}https://play.google.com/store/apps

\bibitem {jua3} J. C. Castro-Palacio, L. Velazquez, J. A. G\'{o}mez-Tejedor, F. J.  Maj\'{o}n and J. A. Monsoriu, Rev. Bras. Ensino Fís. 36 (2014)
2315.

\bibitem {lev} K. Levenberg, Quart. Appl. Math, 2 (1944) 164.

\bibitem {mar} D. Marquardt, SIAM J. Appl Math. 11 (1963) 431.

\bibitem {mau} S. Mau, F. Insulla, E. E. Pickens, Z. Ding and S. C. Dudley, Phys. Teach. 54 (2016) 246.

\end{thebibliography}
\end{document}